# CoDiPhy: A General Framework for Applying Denoising Diffusion Models to the Physical Layer of Wireless Communication Systems

Peyman Neshaastegaran, *Member, IEEE*, and Ming Jian, *Member, IEEE*

*Huawei Technologies Canada Co. Ltd., Ottawa, Ontario, Canada*

peyman.neshaastegaran1@huawei.com

**Abstract**

Generative models, including denoising diffusion models (DM), are gaining attention in wireless applications due to their ability to learn complex data distributions. In this paper, we propose CoDiPhy, a novel framework that leverages conditional denoising diffusion models to address a wide range of wireless physical layer problems. A key challenge of using DM is the need to assume or approximate Gaussian signal models. CoDiPhy addresses this by incorporating a conditional encoder as a guidance mechanism, mapping problem observations to a latent space and removing the Gaussian constraint. By combining conditional encoding, time embedding layers, and a U-Net-based main neural network, CoDiPhy introduces a noise prediction neural network, replacing the conventional approach used in DM. This adaptation enables CoDiPhy to serve as an effective solution for a wide range of detection, estimation, and predistortion tasks. We demonstrate CoDiPhy's adaptability through two case studies: an OFDM receiver for detection and phase noise compensation for estimation. In both cases, CoDiPhy outperforms conventional methods by a significant margin. [1]

## I. INTRODUCTION

The evolution toward 6G networks has been increasingly supported by advancements in artificial intelligence, which offer new approaches to address complex challenges in wireless communication [1]. Building on this progress, generative models such as generative adversarial networks, variational autoencoders, and transformers are emerging as effective solutions for wireless applications, including those at the physical layer (PHY) [2]. Another category of generative models gaining attention in wireless communication is diffusion models (DM) [3]–[12]. Known for their stability and high-quality outputs, DM have shown potential in diverse wireless applications, including network optimization [10], and resource allocation [11]. In this work, we focus specifically on the application of DM at the PHY.

DM have been applied to a range of PHY tasks, including decoding [8], [9], channel estimation [3], [4], signal detection [5]–[7], and synthetic channel generation in the end-to-end learning architecture [13]. While the aforementioned works highlight the potential of DM-based solutions, they exhibit two key limitations. First, these methods often rely on Gaussian signal models to ensure compatibility with the DM framework, particularly in the reverse process, which begins with normal noise. However, this assumption is restrictive, as many wireless PHY problems involve impairments that result in non-Gaussian signal characteristics. Second, current DM-based methods are highly specialized, targeting specific problem sets, such as signal detection under particular wireless architectures or channel estimation in defined scenarios, limiting their broader applicability.

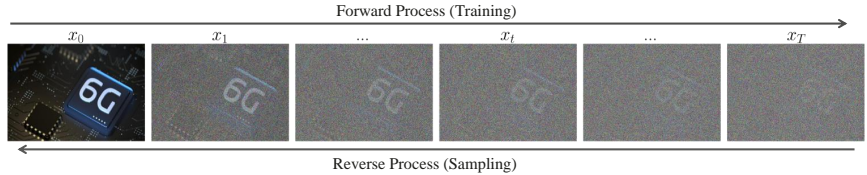

Fig. 1: Diffusion process visualization: the forward process adds noise from $x_0$ to $x_T$, while the reverse process denoises $x_T$ to reconstruct $x_0$.

To address these limitations, in this work we introduce CoDiPhy, a general framework to apply **Co**nditional de-noising **Di**ffusion models in the **Phy**sical layer of wireless communication systems. CoDiPhy employs a conditional encoder to map problem-specific information into a high-dimensional latent space, which serves as a guidance mechanism to direct the diffusion model toward the desired solution. This design allows CoDiPhy to operate with any signal model, overcoming the limitations of prior approaches that rely on Gaussian assumptions. Including the conditional encoder in CoDiPhy is inspired by guidance mechanisms proposed in [14], [15], which employ classifier-free and latent-space guidance to improve flexibility and performance. These mechanisms enable diffusion models to incorporate problem-specific information effectively, achieving a trade-off between fidelity and diversity without retraining or external classifiers. As a proof of concept, CoDiPhy is applied to two distinct tasks: signal detection in an uplink receiver scenario and phase noise (PN) estimation. For signal detection, CoDiPhy achieves a gap of less than 0.5 dB to the linear minimum mean squared error (LMMSE) receiver with perfect channel state information (CSI) at an uncoded bit-error-rate (BER) of $10^{-2}$. For PN estimation, CoDiPhy improves the mean squared error (MSE) over a wide signal-to-noise ratio (SNR) range, achieving up to 6 dB better performance compared to the commonly used pilot symbol assisted modulation (PSAM) estimation.

The rest of this paper is organized as follows. Section II provides a primer on diffusion models Section III details the architecture of the proposed CoDiPhy framework In Section IV, CoDiPhy is adapted to two specific problem sets followed by numerical results. Section V discusses the implications of this work and potential directions for future research.

## II. PRIMER ON DIFFUSION MODELS

DM are generative models that learn complex data distributions through a two-step process: a forward process that corrupts data into noise $q(\mathbf{x}_t|\mathbf{x}_0)$ and a reverse process that reconstructs data by learning $p_\theta(\mathbf{x}_{t-1}|\mathbf{x}_t)$, as shown in Fig. 1. The reverse process uses a model parameterized by $\theta$ to iteratively denoise and recover $\mathbf{x}_0$, minimizing a variational bound between the true and model distributions.

## A. Denoising Diffusion Probabilistic Models

Denoising diffusion probabilistic models (DDPM) use a Markovian process to transform data $\mathbf{x}_0$ into Gaussian noise and learn its reverse to reconstruct $\mathbf{x}_0$. This framework, introduced in [16], comprises two main components: the forward process and the reverse process.

### 1) Forward Process

The forward process incrementally corrupts the data $\mathbf{x}_0$ over $T$ time steps by adding Gaussian noise, transforming the data at each step as[2]:

$$q(\mathbf{x}_t|\mathbf{x}_{t-1}) = \mathcal{N}(\mathbf{x}_t; \sqrt{1-\beta_t}\mathbf{x}_{t-1}, \beta_t\mathbf{I}), \qquad (1)$$

where $\beta_t \in (0,1)$ is a noise variance schedule that controls the magnitude of noise added at each step. By compounding the noise over $T$ steps, the marginal distribution at any time $t$ is given as:

$$q(\mathbf{x}_t|\mathbf{x}_0) = \mathcal{N}(\mathbf{x}_t; \sqrt{\bar{\alpha}_t}\mathbf{x}_0, (1-\bar{\alpha}_t)\mathbf{I}), \qquad (2)$$

where $\alpha_t = 1 - \beta_t$ and $\bar{\alpha}_t = \prod_{s=1}^{t}\alpha_s$. As $T \to \infty$, the data distribution $q(\mathbf{x}_t)$ approaches a standard Gaussian distribution.

### 2) Reverse Process

The reverse process is defined as a Markov chain parameterized by a neural network (NN) $\epsilon_\theta$ to model the conditional distributions $p_\theta(\mathbf{x}_{t-1}|\mathbf{x}_t)$. Starting from standard Gaussian noise $\mathbf{x}_T \sim \mathcal{N}(0,\mathbf{I})$, the reverse process iteratively removes noise from $\mathbf{x}_t$ to recover $\mathbf{x}_0$:

$$p_\theta(\mathbf{x}_{t-1}|\mathbf{x}_t) = \mathcal{N}(\mathbf{x}_{t-1}; \mu_\theta(\mathbf{x}_t, t), \Sigma_\theta(\mathbf{x}_t, t)), \qquad (3)$$

where $\mu_\theta(\mathbf{x}_t, t)$ and $\Sigma_\theta(\mathbf{x}_t, t)$ represent the predicted mean and variance, respectively. Typically, $\Sigma_\theta$ is fixed during training to simplify the model. The mean $\mu_\theta(\mathbf{x}_t, t)$ is computed using the NN output $\epsilon_\theta(\mathbf{x}_t, t)$, which predicts the added noise $\epsilon$ in the data. Specifically, $\mu_\theta$ is formulated as:

$$\mu_\theta(\mathbf{x}_t, t) = \frac{1}{\sqrt{\alpha_t}}\left(\mathbf{x}_t - \frac{\beta_t}{\sqrt{1-\bar{\alpha}_t}}\epsilon_\theta(\mathbf{x}_t, t)\right). \qquad (4)$$

### 3) Training Objective

The training objective for DDPM minimizes the Kullback-Leibler (KL) divergence between the true reverse distribution $q(\mathbf{x}_{t-1}|\mathbf{x}_t, \mathbf{x}_0)$ and the model distribution $p_\theta(\mathbf{x}_{t-1}|\mathbf{x}_t)$. This can be expressed as the following simplified objective:

$$L = \mathbb{E}_{t,\mathbf{x}_0}\left[|\epsilon - \epsilon_\theta(\mathbf{x}_t, t)|^2\right], \qquad (5)$$

where $\epsilon \sim \mathcal{N}(0,\mathbf{I})$.

---

[2] The notation $\mathbf{x} \sim \mathcal{N}(\boldsymbol{\mu}, \boldsymbol{\Sigma})$ indicates that the random vector $\mathbf{x}$ follows a Gaussian distribution with mean vector $\boldsymbol{\mu}$ and covariance matrix $\boldsymbol{\Sigma}$, while $\mathcal{N}(\mathbf{x}; \boldsymbol{\mu}, \boldsymbol{\Sigma})$ represents a Gaussian distribution parameterized by $\mathbf{x}$, with mean vector $\boldsymbol{\mu}$ and covariance matrix $\boldsymbol{\Sigma}$.

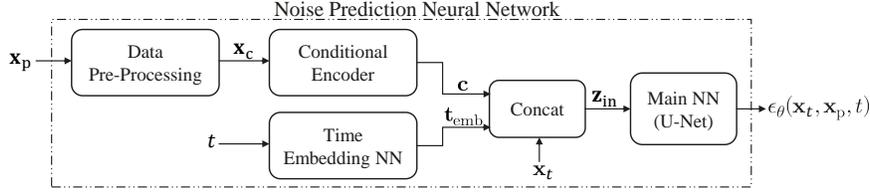

Fig. 2: Block diagram of the NPNN in CoDiPhy, integrating the conditional encoder, time embedding, and main NN.

## B. Denoising Diffusion Implicit Models

Denoising Diffusion Implicit Models (DDIM), introduced in [17], extend the DDPM framework by introducing a deterministic reverse process that accelerates sampling. This formulation retains the same forward process as DDPM, but modifies the reverse process with a tunable parameter $\eta_t$ to control the degree of stochasticity. In DDIM, the reverse process directly computes $\mathbf{x}_{t-1}$ from $\mathbf{x}_t$ using the predicted noise $\epsilon_\theta(\mathbf{x}_t, t)$ and the intermediate predicted data $\hat{\mathbf{x}}_0$:

$$\mathbf{x}_{t-1} = \sqrt{\bar{\alpha}_{t-1}}\hat{\mathbf{x}}_0 + \sqrt{1 - \bar{\alpha}_{t-1} - \eta_t^2}\epsilon_\theta(\mathbf{x}_t, t) + \eta_t \psi, \qquad (6)$$

where $0 \leq \eta_t \leq 1$, and $\psi \sim \mathcal{N}(0, \mathbf{I})$. The intermediate estimate $\hat{\mathbf{x}}_0$ is defined as $\hat{\mathbf{x}}_0 = \frac{\mathbf{x}_t - \sqrt{1-\bar{\alpha}_t}\epsilon_\theta(\mathbf{x}_t, t)}{\sqrt{\bar{\alpha}_t}}$. To further accelerate sampling, DDIM utilizes a subset of time steps $\tau = \{\tau_1, \tau_2, \ldots, \tau_S\}$ from the original $T$ steps used in DDPM, where $\tau[i] > \tau[i-1]$ and $S \ll T$. This subset reduces redundancy in the sampling process while maintaining high fidelity in the generated outputs.

## III. PROPOSED CODIPHY FRAMEWORK

In this section, we introduce our proposed DM-based framework, CoDiPhy, which enables the application of DDPM and DDIM to a wide class of wireless PHY problems. As noted in Section II-A, the reverse diffusion process in standard DDPM starts from normal noise, which can limit its application to wireless problems where the signal model is often non-Gaussian unless approximated or conditioned on additional information, such as known CSI.

To overcome the limitation of the Gaussian assumption in the reverse diffusion process, we introduce a conditional encoder that incorporates all relevant, problem-specific information into the model. This encoder, inspired by the guidance mechanism proposed in [15], broadens the scope of CoDiPhy to accommodate a variety of signal models beyond Gaussian. The encoder extracts these features and generates a high-dimensional encoded condition, denoted as $\mathbf{c}$, which is then integrated into the DDPM framework as a guidance signal. This integration allows CoDiPhy to handle diverse signal models. The following sections provide a detailed description of the noise prediction neural network (NPNN) architecture, as well as the training and sampling processes.

### A. Noise Prediction Network Architecture

The architecture of the NPNN, the core of the CoDiPhy framework, consists of three key components: a conditional encoder, a time embedding layer, and the main NN for noise prediction. As shown in Fig. 2, these components collaborate to enable CoDiPhy's functionality. Each network and its role within the architecture are outlined below.

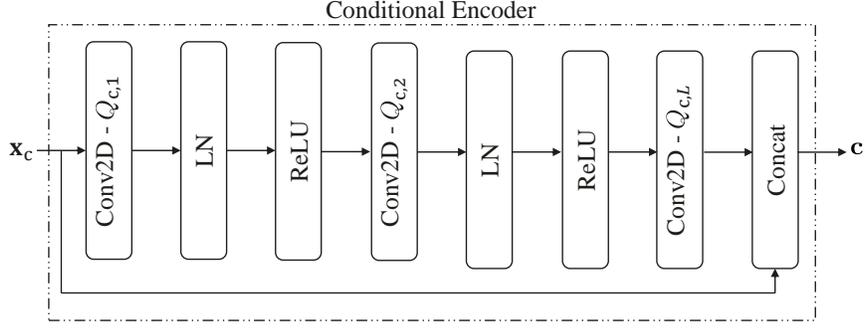

Fig. 3: Conditional encoder architecture, transforming problem-specific data into the encoded condition $\mathbf{c}$.

*1) Conditional Encoder*

The conditional encoder maps problem-specific observations $\mathbf{x}_\mathrm{p}$ into a high-dimensional space, enabling the main NN to guide the model toward the desired output. As shown in Fig. 3, the encoder consists of three 2D convolutional layers, where the first two are followed by layer normalization and a ReLU activation, and the third produces the output. To ensure dimensionality matching, a pre-processing stage converts the input into $\mathbf{x}_\mathrm{c}$ with dimensions $D_h \times D_w \times D_{ch}$ (excluding the batch size). The convolutional layers use numbers of kernels $Q_{\mathrm{c},1}$, $Q_{\mathrm{c},2}$, and $Q_{\mathrm{c},L}$, with $Q_{\mathrm{c},L}$ defining the latent space dimension. Padding is set to "same" to ensure that $D_h$ and $D_w$ remain unchanged throughout the layers. Setting $Q_{\mathrm{c},L} > D_{ch}$ allows the encoder to capture richer features. To retain low-level input features while incorporating higher-level abstractions, a skip connection concatenates the final convolution's output with the original input $\mathbf{x}_\mathrm{c}$ to obtain the resulting encoded condition $\mathbf{c} = f_{\mathrm{ConEnc}}(\mathbf{x}_\mathrm{p})$.

*2) Time Embedding Layer*

The time embedding layer encodes the time step $t$ from the diffusion process into a high-dimensional representation, $\mathbf{t}_{\mathrm{emb}}$, with a tunable dimension $Q_t$, which corresponds to the output size of the dense layer used. By transforming the discrete time step into a continuous representation, the network captures temporal dependencies, enhancing its ability to learn reverse process dynamics.

*3) Main NN*

The main NN in CoDiPhy, shown in Fig.2, is based on the U-Net architecture, a widely used model in DM literature. Originally proposed for image segmentation, the U-Net effectively processes hierarchical information and captures fine-grained details. For more on U-Net, readers can refer to [18].

In CoDiPhy, the input to the main NN at time step $t$, denoted as $\mathbf{z}_{\mathrm{in}}$, combines the high-dimensional condition $\mathbf{c}$ from the conditional encoder, the embedded time $\mathbf{t}_{\mathrm{emb}}$, and either the ground truth (during training) or the noisy input (during sampling), $\mathbf{x}_t$. This input is expressed as $\mathbf{z}_{\mathrm{in}} = \mathrm{concat}(\mathbf{c}, \mathbf{t}_{\mathrm{emb}}, \mathbf{x}_t)$. Using this input, the main NN predicts the noise, $\epsilon_\theta(\mathbf{x}_t, \mathbf{x}_p, t)$, calculated as $\epsilon_\theta(\mathbf{x}_t, \mathbf{x}_p, t) = f_{\mathrm{MainNN}}(\mathbf{z}_{\mathrm{in}})$.

## B. Training Procedure for CoDiPhy

The training of CoDiPhy builds on the standard DDPM framework, incorporating problem-specific conditioning via the conditional encoder. For training, $\mathbf{x}_t$ is computed as:

$$\mathbf{x}_t = \sqrt{\bar{\alpha}_t}\mathbf{x}_0 + \sqrt{1 - \bar{\alpha}_t}\epsilon, \tag{7}$$

**Algorithm 1** Training Algorithm for CoDiPhy
─────────────────────────────────────────────
**Input:** NN Hyper-Parameters, Time Steps $T$, Noise Schedule $\beta_t$, Problem Dataset $\mathcal{S}$
**Output:** Trained NPNN $\epsilon_\theta(\cdot)$

1: **while** not converged **do**
2:     Sample $t \sim \mathrm{Uniform}\{1,\ldots,T\}$
3:     Sample $(\mathbf{x}_p, \mathbf{x}_0)$ from $\mathcal{S}$
4:     Sample $\epsilon \sim \mathcal{N}(0, \mathbf{I})$
5:     Pass $\mathbf{x}_p$ to conditional encoder to obtain $\mathbf{c}$
6:     Pass $t$ to time embedding layer to obtain $\mathbf{t}_{\mathrm{emb}}$
7:     Calculate $\mathbf{x}_t$ using (7)
8:     Let $\mathbf{z}_{\mathrm{in}} = \mathrm{concat}(\mathbf{c}, \mathbf{t}_{\mathrm{emb}}, \mathbf{x}_t)$
9:     Pass $\mathbf{z}_{\mathrm{in}}$ to main NN to obtain $\epsilon_\theta(\mathbf{x}_t, \mathbf{x}_p, t)$
10:    Take gradient descent step on $\nabla_\theta \left[ \|\epsilon_\theta(\mathbf{x}_t, \mathbf{x}_p, t) - \epsilon\|^2 \right]$
11: **end while**
12: **return** $\epsilon_\theta(\cdot)$
─────────────────────────────────────────────

where $\epsilon \sim \mathcal{N}(0, \mathbf{I})$, and $\mathbf{x}_0$ represents the desired solution or ground truth. The training objective minimizes the difference between the predicted and true noise by optimizing a loss function derived from the evidence lower bound [15], [16]:

$$\mathcal{L}_\theta = \mathbb{E}_{\mathbf{x}_t, \mathbf{x}_p, t} \left[ \|\epsilon_\theta(\mathbf{x}_t, \mathbf{x}_p, t) - \epsilon\|^2 \right] \tag{8}$$

Algorithm 1 outlines the CoDiPhy training steps.

*C. Sampling Procedure for CoDiPhy*

In standard DDPM sampling, the model denoises input over $T$ steps, starting from Gaussian noise $\mathbf{x}_T$ to generate $\mathbf{x}_0$. Most existing works applying DM to PHY problems replace $\mathbf{x}_T$ with a corrupted signal, assuming a Gaussian signal model. CoDiPhy avoids this by passing all relevant observations, such as the corrupted signal, to the conditional encoder and starting the sampling process with $\mathbf{x}_T \sim \mathcal{N}(0, \mathbf{I})$, fully aligned with DDPM. To accelerate sampling, CoDiPhy uses the updating rule in (6) from [17], reducing iterations from $T$ to $S$ with the sampling set $\tau$ introduced in Section II-B. Algorithm 2 summarizes the CoDiPhy sampling steps.

### IV. ADAPTATION TO SPECIFIC PROBLEM SETS

As discussed, CoDiPhy is designed for adaptability across diverse PHY problems. This section outlines the key steps for adaptation, followed by two case studies showcasing its application and comparisons with conventional methods.

The first step is to define the input to the conditional encoder, $\mathbf{x}_p$, which includes observation signals containing essential information. DM excel at learning complex data distributions, so it is crucial to provide informative data while avoiding redundancy. For instance, in channel estimation, providing transmitted and received pilots enables the model to estimate the SNR, rendering additional information like AWGN power unnecessary. The second step is

**Algorithm 2** Sampling Algorithm for CoDiPhy

**Input:** Sampling Set $\tau$, Noise Schedule $\beta_t$, Parameter $\eta_t$, Trained NPNN $\epsilon_\theta(\cdot)$, Problem-Specific Data $\mathbf{x}_p$

**Output:** Generated Solution $\mathbf{x}_0$

1: Pass $\mathbf{x}_p$ to conditional encoder to obtain $\mathbf{c}$
2: Sample $\mathbf{x}_{\tau_S} \sim \mathcal{N}(0, \mathbf{I})$
3: **for** $i = S$ to $1$ **do**
4:     Pass $\tau_i$ to time embedding layer to obtain $\mathbf{t}_{\text{emb}}$
5:     Let $\mathbf{z}_{\text{in}} = \text{concat}(\mathbf{c}, \mathbf{t}_{\text{emb}}, \mathbf{x}_{\tau_i})$
6:     Pass $\mathbf{z}_{\text{in}}$ to main NN to obtain $\epsilon_\theta(\mathbf{x}_{\tau_i}, \mathbf{x}_p, \tau_i)$
7:     $\psi \leftarrow \mathcal{N}(0, \mathbf{I})$ if $i > 1$, else $\psi \leftarrow 0$
8:     Update the denoised sample according to
$$\mathbf{x}_{\tau_{i-1}} = \sqrt{\bar{\alpha}_{\tau_{i-1}}} \left( \frac{\mathbf{x}_{\tau_i} - \sqrt{1 - \bar{\alpha}_{\tau_i}} \epsilon_\theta(\mathbf{x}_{\tau_i}, \mathbf{x}_p, \tau_i)}{\sqrt{\bar{\alpha}_{\tau_i}}} \right) \\ + \sqrt{1 - \bar{\alpha}_{\tau_{i-1}} - \eta_{\tau_i}^2} \epsilon_\theta(\mathbf{x}_{\tau_i}, \mathbf{x}_p, \tau_i) + \eta_{\tau_i} \psi.$$
9: **end for**
10: $\mathbf{x}_0 \leftarrow \mathbf{x}_{\tau_0}$
11: **return** $\mathbf{x}_0$

to define the ground truth, representing the ideal solution. While a straightforward approach is to use true solution values during training, our trials indicate that it is often more effective to generate the ground truth using known methods under ideal scenarios. For example, in a beamforming design problem, one might use a conventional method assuming perfect CSI and high SNR to create the ground truth. This allows the model to replicate responses without requiring ideal conditions during sampling.

Another critical aspect is selecting appropriate hyperparameters. Increasing U-Net kernels enhances the model's capacity to capture complex relationships, while excessively enlarging the conditional encoder adds little benefit. The learning rate is set around $1 \times 10^{-4}$, as in [16], and a sigmoid noise variance schedule [17] ensures granular coverage from 0 to 1. Algorithm 3 outlines the steps for adapting CoDiPhy to specific PHY problems.

*A. Case Study 1: OFDM Signal Detection in Uplink*

In this case study, we consider an OFDM uplink scenario where a single-antenna UE transmits to an 8-antenna base station over a tapped delay line (TDL) channel model, consistent with 3GPP standards. Key parameters for the setup are provided in Table I. To adapt CoDiPhy as the receiver, the following settings are applied per Algorithm 3:

- **Problem-Specific Data** ($\mathbf{x}_p$): The primary input is the post-FFT received time-frequency resource grid. Additionally, a resource grid containing only pilot symbols (with data locations set to zero) is included.
- **Desired Solution** ($\mathbf{x}_0$): While transmitted signals are a straightforward choice, we find that using the LMMSE equalizer output under ideal conditions (perfect CSI) provides a more effective training target. This approach not only better captures the solution distribution but also offers improved soft information to the decoder, as reflected in the BER results.

**Algorithm 3** Adaptation of CoDiPhy to PHY Problems
1: $\mathbf{x}_p \leftarrow$ Noisy observations, Pilot signals, etc...
2: $\mathbf{x}_0 \leftarrow$ Desired solution or the ground truth
3: Select NPNN hyper-parameters, $\beta_t$, $\eta_t$, and $T$
4: Train the NPNN using Algorithm 1.
5: Generate the solution using Algorithm 2.

| Parameter | Value |
|---|---|
| Carrier Frequency | 2.5 GHz |
| Channel Model | TDL-D (Block Fading) |
| Maximum Delay Spread | 100 ns |
| OFDM Symbols per Frame | 14 |
| FFT size/ CP Length / Subcarrier Spacing | 64/ 6 / 30 kHz |
| Signal Constellation | 16-QAM |
| Pilot Symbols per Frame / Pilot Constellation | 2 / QPSK |
| FEC / Block Length / Code Rate | LDPC / 1538 / 7/8 |
| $Q_{c,1}/Q_{c,2}/Q_{c,L}$/ Kernel Size / $Q_t$ | 64/64/128/(3,3) / 16 |
| Learning Rate / Training Epochs | $8 \times 10^{-5}$ / 10000 |
| $\eta_t$ / $T$ / $S$ | 1.0 / 500 / 15 |
| $\min(\beta_t)/\max(\beta_t)$ | $5 \times 10^{-4}$ / $10^{-2}$ |

TABLE I: The table lists the parameters for Case Study 1, with DM-related parameters also reused in Case Study 2.

During training, a new resource grid is generated at each iteration, sampling from the channel and AWGN distributions to create received signals. The training and sampling phases are implemented according to Algorithms 1 and 2.

To evaluate CoDiPhy, we compare its performance with two baseline methods: the LMMSE equalizer with perfect CSI (PCSI) and the LMMSE equalizer with least squares-based imperfect CSI (ICSI). Unlike these baselines, CoDiPhy requires no CSI information during training or sampling. Fig. 4 shows color maps of normalized residual error between transmitted and estimated resource grids, with MSE values over an SNR range of -4 to 5 dB displayed atop each map. Six maps depict CoDiPhy's reverse diffusion steps. While LMMSE-ICSI suffers over 4 dB MSE degradation due to imperfect CSI, CoDiPhy achieves MSE within 0.1 dB of LMMSE-PCSI without any CSI.

Uncoded and coded BER results are shown in Fig. 5, comparing CoDiPhy with two ground truth options: LMMSE output (GT1) and transmitted signals (GT2). CoDiPhy with GT1 performs within 0.5 dB of PCSI over all SNR ranges, demonstrating its capability as an OFDM receiver. In contrast, GT2 exhibits weaker coded BER performance. Tests with 64-QAM and 256-QAM constellations confirm these findings, with CoDiPhy-GT1 staying within 1 dB of PCSI at $10^{-2}$ uncoded BER and $10^{-6}$ coded BER. The gap between PCSI and ICSI increases with higher constellation sizes, though these results are omitted due to space constraints.

*B. Case Study 2: Phase Noise Estimation*

For the second case study, we examine PN estimation in single-carrier systems with high-order modulations, common in wireless backhaul. The post-equalization model is $y = se^{j\phi} + n$, where $y$ is the received signal, $s$ the transmitted signal, $\phi$ the PN, and $n$ the AWGN. The conventional PSAM approach embeds one pilot every $P - 1$ data symbols and interpolates PN between pilot positions. However, PSAM performance is fundamentally constrained by the Nyquist sampling theorem. For this study, we use $P = 50$, a PN level of -88 dBc/Hz at 100 kHz offset, and 256-QAM modulation. To enhance PSAM, CoDiPhy estimates PN by processing pilot sections, each containing $P$ symbols (one pilot and $P - 1$ data symbols). The adaptation uses the following parameters:

- **Problem-Specific Data** ($\mathbf{x}_p$): To capture the time-varying nature of PN, we provide $y/s_{\text{PSAM}}$, where $s_{\text{PSAM}}$ is the hard decision made based on the PSAM estimate. The AWGN power is also included to help distinguish between PN and additive noise contributions.
- **Desired Solution** ($\mathbf{x}_0$): The true PN values are chosen as the desired solution, as they represent the estimation target and are sufficient for effective training in this task.

Using this parameter set, we follow Algorithm 3 to train and apply CoDiPhy to the PN estimation task. Fig.6 shows the evolution of PN estimates during sampling, with the MSE of PSAM ($\text{MSE}_P$) and CoDiPhy ($\text{MSE}_C$) displayed above each subplot. Here, the real part of $e^{j\phi_{\text{est}}}$ from CoDiPhy is compared against PSAM for a random pilot section. Initially, CoDiPhy's MSE is higher than PSAM's but improves with each step, eventually surpassing it, demonstrating CoDiPhy's iterative refinement capability.

To assess the generalizability of CoDiPhy, we evaluate its performance across various SNR and PN levels, comparing it with PSAM. Fig. 7 demonstrates that CoDiPhy consistently outperforms PSAM, achieving significantly lower MSE values. At high SNRs, the improvement over PSAM reaches 6 dB, showcasing the model's ability to effectively extract PN estimates using the provided conditional information.

V. DISCUSSION AND FUTURE WORK

In this paper, we introduced CoDiPhy as a versatile framework for addressing a variety of wireless PHY problems. We demonstrated its adaptability and effectiveness through two case studies, focusing on detection and estimation tasks. While results for additional applications, such as PAPR reduction and OFDM receivers with time-varying channels, have also been obtained, they are not included here due to space constraints.

A detailed complexity analysis was not provided in this work, as the computational complexity of CoDiPhy is highly problem-dependent. For instance, problems involving high-dimensional inputs require a more complex conditional encoder with additional convolutional layers, larger kernels, and more feature maps to effectively extract the necessary features. Similarly, solving problems with highly intricate data distributions may necessitate augmenting the U-Net in the main NN, leading to increased computational demands. As such, a meaningful complexity analysis for CoDiPhy must be conducted on a case-by-case basis.

For future work, we plan to evaluate the generalizability of CoDiPhy under out-of-distribution scenarios to assess its robustness. We will also compare its performance with existing discriminative models in specific tasks to highlight the advantages of generative approaches in terms of learning capabilities. Additionally, we aim to explore non-traditional applications of CoDiPhy, such as beamforming in reconfigurable intelligent surfaces and signal detection in free-space optical communications impacted by non-linearities.

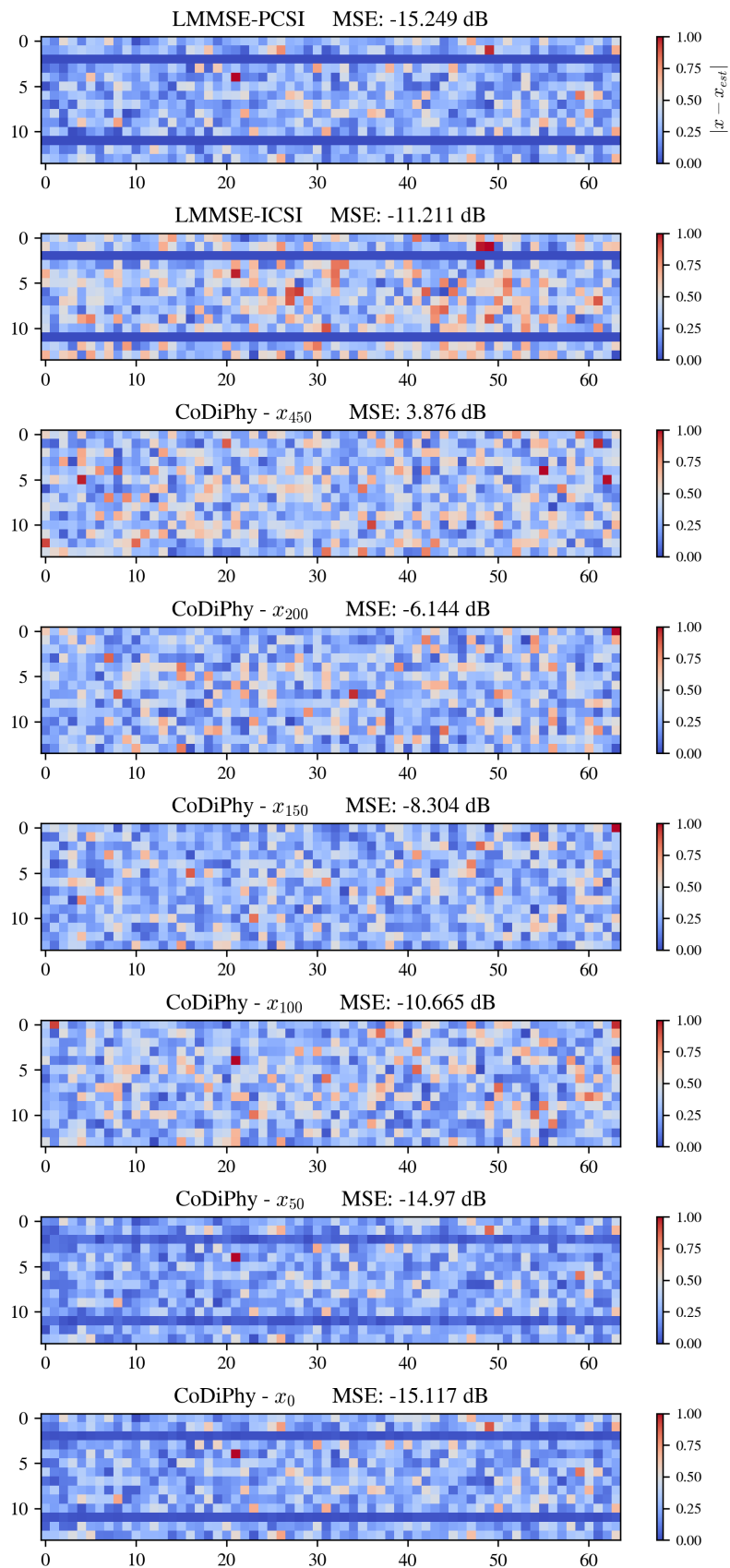

Fig. 4: Color maps of residual error for various schemes, including six steps of CoDiPhy's reverse diffusion. MSE values for SNRs from -4 to 5 dB are shown above each map.

March 11, 2025 DRAFT

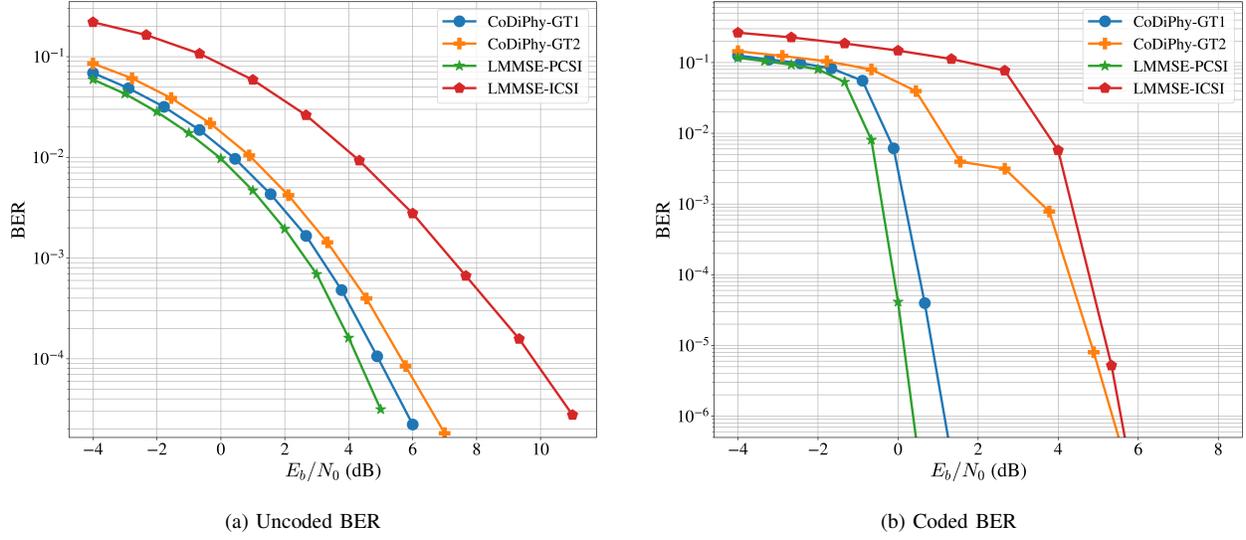

(a) Uncoded BER

(b) Coded BER

Fig. 5: Performance comparison of CoDiPhy with conventional schemes, showing uncoded BER and coded BER across various SNR levels for different ground truth options.

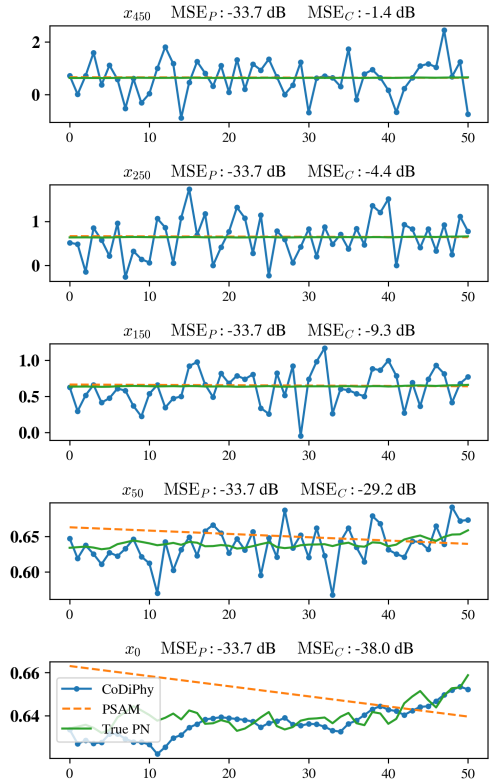

Fig. 6: Evolution of PN estimates during the reverse diffusion process, showing $\text{Re}\{e^{j\phi_{\text{est}}}\}$ in radians. Each step illustrates the refinement of PN estimation toward the true values.

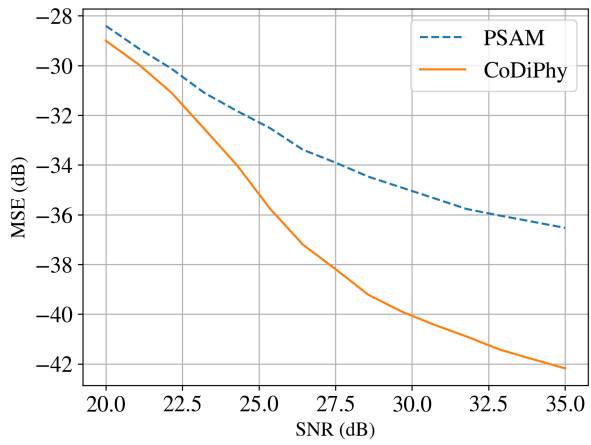 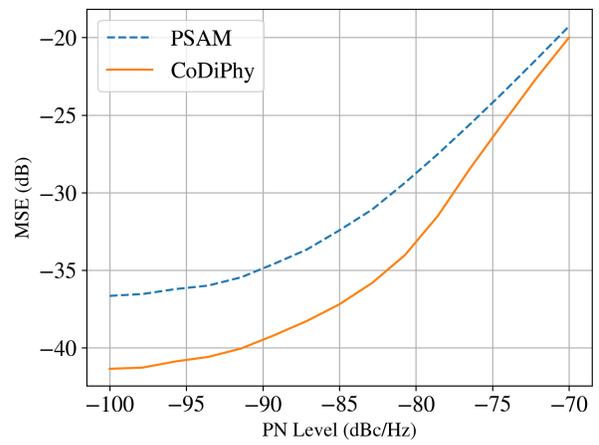

(a) MSE across SNR levels

(b) MSE across PN levels

Fig. 7: PN estimation performance comparison across SNR and PN levels, highlighting the MSE of CoDiPhy and PSAM.